\documentclass[aps,prl,twocolumn]{revtex4-2}
\usepackage{braket}

\usepackage{graphicx}
\usepackage{xcolor}

\usepackage{hyperref}
\hypersetup{colorlinks=true, linkcolor=blue, citecolor=blue, urlcolor=blue}

\newcommand{\sect}[1]{\vspace{0.3em}{\it #1.}---}

\newcommand{\eref}[1]{Eq.~(\ref{#1})}

\newcommand{\fref}[1]{Fig.~\ref{#1}}
\newcommand{\frefs}[1]{Figs.~\ref{#1}}
\newcommand{\rcite}[1]{Ref.~\cite{#1}}
\newcommand{\rcites}[1]{Refs.~\cite{#1}}

\begin{document}

\title{Work-minimizing protocols in driven-dissipative quantum systems: \\ An impulse-ansatz approach}

\date{\today}


\author{Masaaki Tokieda}
\email{tokieda.masaaki.b01@kyoto-u.jp}
\affiliation{Department of Chemistry, Graduate School of Science, Kyoto University, Kyoto, Japan}

\begin{abstract}
The second law of thermodynamics sets a lower bound on the work required to drive a system between thermal equilibrium states, with equality attained in the quasistatic limit.
For finite-time processes, work in excess of this bound is inevitably dissipated, motivating the search for driving protocols that minimize this excess work.
While classical stochastic systems have been extensively explored, quantum analyses remain limited and often rely on Markovian master equations valid only in the weak-coupling regime.
Here, we study minimal work protocols for representative two-level systems coupled to a harmonic-oscillator bath using a numerically exact method.
Inspired by known optimal solutions for Brownian oscillators, we introduce an impulse ansatz that incorporates possible boundary impulses and test it across a wide range of bath parameters.
We find that impulse-like features remain nearly optimal in the quantum, non-Markovian regime at short times.
We also identify cases in which the widely used Markovian master equation fails even at weak coupling, highlighting the necessity of going beyond standard approximations in the quantum regime for finite-time thermodynamic optimization.
\end{abstract}

\maketitle

\sect{Introduction}
Thermodynamics of small classical systems has been intensively investigated in the past decade, motivated by precise experiments on colloidal particles and microscopic engines \cite{Blickle12,Berut12,Martinez16,Ciliberto17,Barros24}.
As experiments push to smaller scales and more controllable settings, regimes emerge where thermal and quantum fluctuations coexist, and classical descriptions become insufficient.
This naturally motivates the field of quantum thermodynamics \cite{Campbell25}, with recent experimental progress in systems such as ultracold atoms \cite{Bouton21}, nitrogen vacancy centers in diamond \cite{Klatzow19}, and nuclear magnetic resonance
\cite{Peterson19}.

Various formulations of quantum thermodynamics have been developed within the framework of open quantum dynamics.
The standard Markovian approach \cite{Spohn78,SL78,Alicki79,AK18}, based on the Gorini-Kossakowski-Sudarshan-Lindblad (GKSL) master equation \cite{GKS,Lindblad}, defines thermodynamic quantities solely through the system degrees of freedom and is primarily applicable in the weak system-bath coupling regime.
To achieve a consistent description irrespective of the coupling strength, several approaches grounded in a full system-bath Hamiltonian have been proposed: Those relying only on the system degrees of freedom \cite{Hilt11,Rivas20,Colla22}, those defining entropy through the build-up of system-bath correlations \cite{ELB10,Landi21}, and those tailored to the slowly driven regime \cite{Dou18}.
Recently, Koyanagi and Tanimura introduced an alternative formulation \cite{KT24_1} that expresses the second law for general processes,
including temperature-varying ones, in terms of a lower bound on a dimensionless work defined via the time variation of the Hamiltonian weighted by the inverse temperature.
In addition, their framework recovers textbook thermodynamic relations through entropic potentials \cite{PV02}, making explicit the Legendre-transform structure linking intensive and extensive variables, and offers a consistent description applicable to both classical and quantum regimes.

This work is motivated by the extension of the framework to nonequilibrium regimes \cite{KT24_2}, where the dimensionless work remains bounded from below and defines the corresponding entropic potentials through its minimum value.
This raises the question of which driving protocol achieves this minimum.
This question was not addressed in \rcite{KT24_2}, which relied on brute-force numerical optimization.
Here, we seek deeper insight into the minimal-work protocol by analyzing simpler, representative examples.

The optimization of thermodynamic processes in nonequilibrium regimes has long been studied under the label of finite-time thermodynamics.
Motivations include maximizing power output in macroscopic \cite{Curzon75,Andresen84}, microscopic \cite{Geva92,Rezek09,Brandner20,Ye22,Frim22}, and information heat engines \cite{Zhou24}, minimizing the average and fluctuation of excess work to improve free-energy estimates \cite{SS07,MSS08,Geiger10}, and minimizing entropy production to approach finite-time Landauer bounds \cite{Miller20,LLKP22,VS22,RL23}.
Remarkably, for Brownian oscillator systems, \rcites{SS07,MSS08} showed that the minimal-work path features discontinuities at the beginning and end times:
A finite jump in overdamped systems, which is ubiquitous in the short-duration limit \cite{Blaber21} and for which recent experimental results provide compelling evidence \cite{Loos24}, and delta-functional impulses for underdamped systems.
Inspired by these findings, we introduce an ansatz that captures possible boundary impulses and test it on prototypical two-level systems.
Using the bath oscillator model, we explore a wide range of bath parameters via a numerically exact method.
We find that the ansatz is particularly effective in the short-time region, indicating the relevance of impulse-like features in the quantum domain.
Lastly, we assess the standard Markovian approach by treating it as an approximation to the full system-bath description and show that it fails even in the weak-coupling regime.

\sect{Problem settings}
We set $\hbar = 1$.
Consider a bath oscillator model
$H(\lambda(t)) = H_S(\lambda(t)) + H_B + V_S X_B$,
where $H_S(\lambda(t))$ is a system Hamiltonian with a control field $\lambda(t)$, $H_B = \sum_m \omega_m b_m^\dagger b_m$ is a bath Hamiltonian with mode frequencies $\omega_m$ and the bosonic annihilation (creation) operators $b_m$ ($b_m^\dagger$), and $X_B = \sum_m c_m (b_m+b_m^\dagger)$ couples to the system operator $V_S$ with strengths $c_m$ \cite{FV63,Ullersma66,CL83model,FLO88,Weiss08}.
The total state $\rho(t)$ evolves as $\dot{\rho}(t) = - i [H(\lambda(t)), \rho(t)]$.
The reduced system state, $\rho_S(t) = {\rm tr}_B[\rho(t)]$ with ${\rm tr}_B$ the trace over the bath, is uniquely determined, for a thermal bath at inverse temperature $\beta$, by the bath correlation function $L(t) = (1/\pi) \int_{-\infty}^\infty d\omega \, J(\omega) / (1 - e^{-\beta \omega})$, where the spectral density $J(\omega) = \pi \sum_m c_m^2 [\delta(\omega-\omega_m) - \delta(\omega+\omega_m)]$
encodes the bath modes and coupling constants.
To model irreversible dynamics, we consider smooth spectral densities $J(\omega)$ corresponding to a continuous distribution of bath modes.

Numerous studies indicate that, when the system steady state is unique, it coincides with the reduced Gibbs state, $\lim_{t \to \infty} \rho_S(t) = {\rm tr}_B[e^{-\beta H(\lambda)}]/{\rm tr}[e^{-\beta H(\lambda)}]$, where ${\rm tr}$ denotes the trace over the total space, for a fixed $\lambda(t) = \lambda$ \cite{GSI88,Tanimura14,TMCA22}, although a general proof remains open.
Assuming this holds in general, the model offers a dynamical description of thermalization, thereby justifying its application to finite-time thermodynamic processes \cite{KT24_1,KT24_2}.

\begin{figure}[t]
\includegraphics[keepaspectratio, scale=0.25]{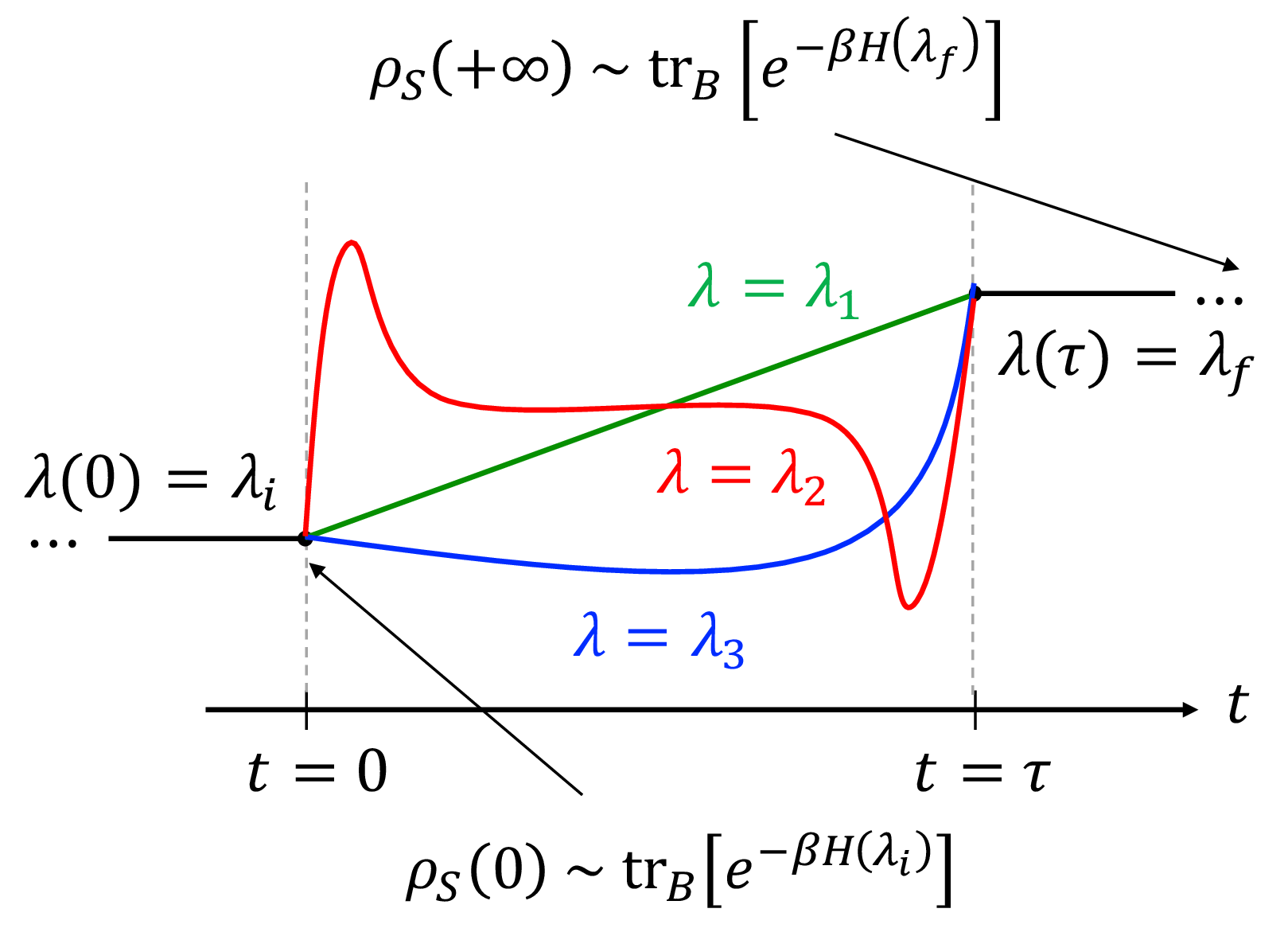}
\caption{
Schematic illustration of the problem setting.
The system is driven between thermal equilibrium states at $\lambda_i$ and $\lambda_f$ by a control field $\lambda(t)$ varied only during a finite interval $\tau$.
The work depends on the protocol $\{\lambda(t)\}_{0\le t\le \tau}$, and our goal is to identify the protocol that minimizes it.
}
\label{fig:settings}
\end{figure}

We focus on isothermal processes that drive the system from the thermal state at $\lambda_i = \lambda(0)$ to that at $\lambda_f = \lambda(t\ge\tau)$, where the control field $\lambda(t)$ varies only during a finite window $0 \le t \le \tau$, as illustrated in \fref{fig:settings}.
The work required to drive the system is given as a functional of $\{ \lambda(t) \}_{0 \leq t \leq \tau}$ by $W_\tau[\lambda]
= \int_0^\tau dt\, {\rm tr}_S\!\left[\dot H_S(\lambda(t))\,\rho_S(t)\right]$,
with ${\rm tr}_S$ the trace over the system.
For the present Hamiltonian model, the work is expected to satisfy \cite{Tasaki02,dscr.vs.cont}
\begin{equation}
    W_\tau [\lambda] \geq \Delta F
    \label{eq:settings_2nd.law}
\end{equation}
where $\Delta F = F_f - F_i$ and $F_{i/f}=-(1/\beta)\ln{\rm tr}[e^{-\beta H(\lambda_{i/f})}]$.
Since $\Delta F$ corresponds to the equilibrium free-energy difference, this inequality can be interpreted as the second law of thermodynamics.

The equality in \eref{eq:settings_2nd.law} is achieved in the quasistatic limit $\tau \to \infty$ \cite{ST20,KT22_1}.
For finite $\tau$, the work exceeds $\Delta F$ and depends on the chosen protocol $\lambda(t)$.
Our aim in this Letter is to identify, or in practice closely approximate, the optimal protocol $\lambda^*(t)$ that minimizes $W_\tau[\lambda]$.
Most previous studies addressed classical Langevin systems in the overdamped limit.
In this connection, we recall that the classical limit of the bath oscillator model with an Ohmic spectral density $J(\omega)\propto\omega$ reproduces the Markovian Langevin equation \cite{CL83FP}, from which the overdamped limit follows.
Thus, the exact quantum treatment of the model naturally extends these studies to underdamped, quantum, and non-Markovian regimes.

\sect{Impulse ansatz}
Previous studies have obtained the optimal protocol $\lambda^* (t)$ through a variety of approaches, including analytic solutions of the variational equations \cite{SS07,MSS08,EKLB10}, brute-force numerical optimization \cite{TE08,Geiger10,SH18,Engel23}, optimal transport  \cite{AMG11,AGMG12,Nakazato21,VS23,Kamijima25,Nagase25}, and thermodynamic geometry \cite{Sivak12,Bonanca14,Wadia22}.
Recently, \rcite{Zhong24} established the equivalence between the optimal-transport and thermodynamic-geometry frameworks in the slowly driven regime and used this insight to propose an informed ansatz for $\lambda^* (t)$, which circumvents the computational cost of optimal transport.

\begin{figure}[t]
\includegraphics[keepaspectratio, scale=0.25]{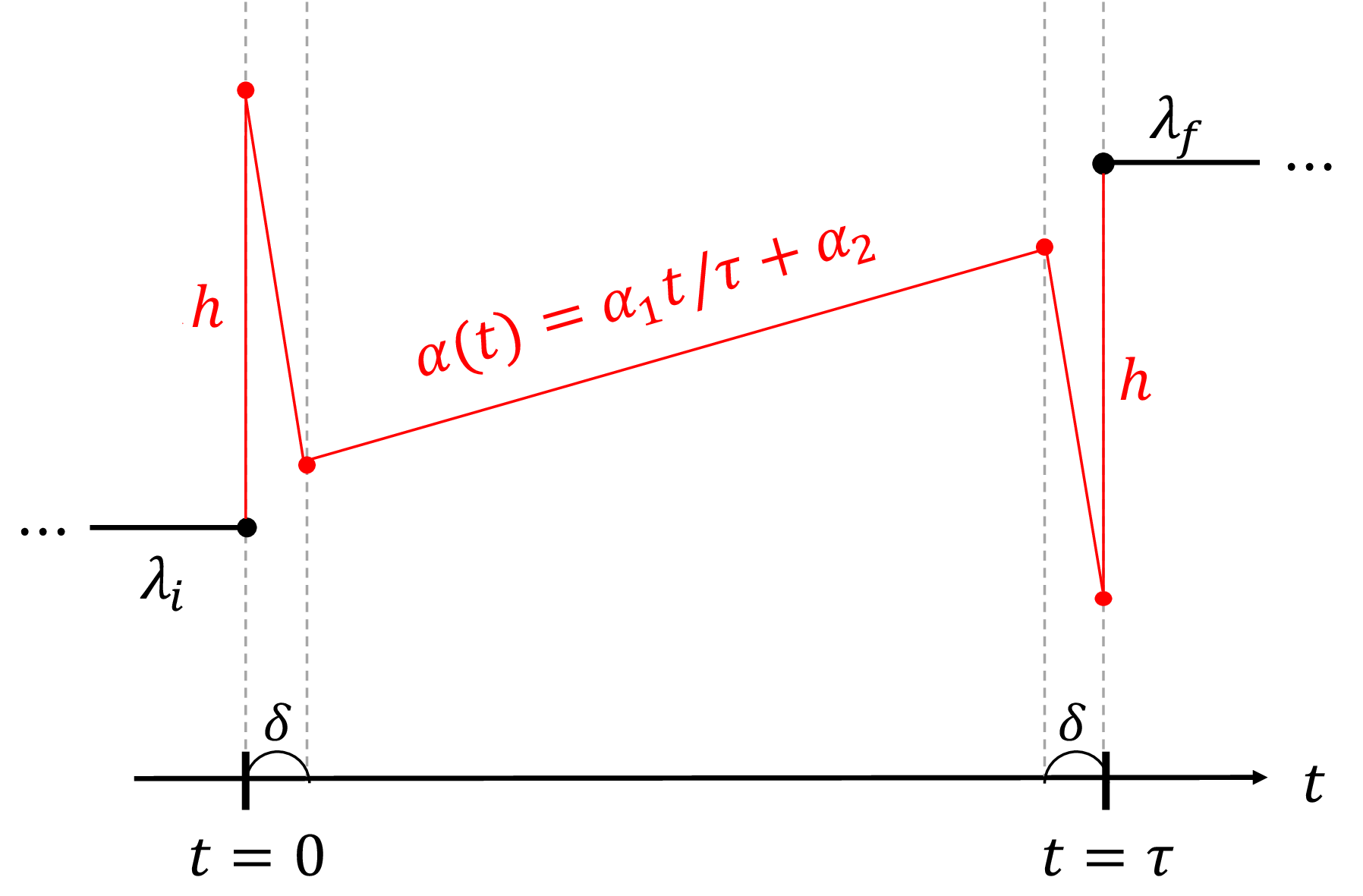}
\caption{
Impulse ansatz characterized by three parameters $(h,\alpha_1,\alpha_2)$, with a fixed impulse width $\delta$ treated as a hyperparameter.
}
\label{fig:ansatz}
\end{figure}

Here, we adopt a similar strategy based on an informed ansatz, but with parameters determined numerically by minimizing the work $W_\tau[\lambda]$. Our ansatz is motivated by previous findings that discontinuities in the control field can reduce the work. A paradigmatic case is the classical Brownian oscillator, for which the analytic optimal protocol exhibits delta-functional impulses at the beginning and end times in the underdamped regime \cite{MSS08}.
Inspired by this structure, we consider an impulse ansatz in which $\lambda(t)$ is allowed to change abruptly near $t=0$ and $t=\tau$, with a smooth interpolation in between. Guided by the analytic solution for a Brownian particle in a moving harmonic trap (reviewed in the Supplemental Material \cite{SM}), we assume linear interpolation and symmetric boundary impulses, as illustrated in \fref{fig:ansatz}.
Instead of ideal delta functions, which are experimentally unrealistic and lead to ambiguities in the numerical evaluation of $W_\tau[\lambda]$, we introduce a finite impulse width $\delta$, interpreted as the minimal time resolution and kept fixed.
The resulting protocol is specified by three parameters $(h,\alpha_1,\alpha_2)$, which we refer to as the three-parameter impulse ansatz (IMP3).

\sect{Results and discussions}
We demonstrate the utility of IMP3 using two prototypical two-level systems.
We adopt the Drude spectral density
$J(\omega)=\gamma^2 \xi \omega/(\omega^{2}+\gamma^{2})$ \cite{OhmicAnomaly}, where $\gamma$ sets the bath memory time and $\xi$ the system-bath coupling strength.
This form yields an exponentially decaying friction kernel, $\propto \gamma \xi e^{-\gamma t}$, observed in viscoelastic media \cite{Loos24}.
To demonstrate the broad applicability of IMP3,  we survey 18 bath parameter sets with $\gamma=0.2,1,5$, $\beta=0.2,1,5$, and $\xi=0.2,1$, in units where the level spacing of $H_S(\lambda_i)$ is unity. We consider operation times $0.5 \le \tau \le 15$.
The impulse width is fixed at $\delta = 10^{-2}$, corresponding to controls two orders of magnitude faster than the system timescale, a regime attainable with the latest experimental techniques \cite{Rieser_private}.
For each parameter set, the work $W_\tau[\lambda]$ and free-energy difference $\Delta F$ are evaluated using the method of hierarchical equations of motion (HEOM) \cite{TK89,YYLS04,IT05,IF09,MT25}.
Work minimization is carried out numerically using a derivative-based iterative algorithm.
Further computational details are provided in the Supplemental Material \cite{SM}. Given the parameterized ansatz for $\lambda(t)$, we minimize $W_\tau[\lambda]$ and denote the resulting optimal protocol by $\lambda^*(t)$.

We assess the performance of IMP3 by comparing it with three reference protocols.
(i) The first is the naive linear protocol $\lambda_{\rm linear}(t) = (\lambda_f-\lambda_i) \, t/\tau + \lambda_i $.
(ii) Second, to test the efficiency, we consider a three-parameter polynomial ansatz (POLY3)
$\lambda_{\rm POLY3}(t) = \lambda_{\rm linear}(t) + t/\tau \, (t/\tau-1)(\alpha_1 (t/\tau)^2 + \alpha_2 t/\tau + \alpha_3)$, which aligns with the boundary conditions, and expect $W_\tau[\lambda^*_{\rm IMP3}] < W_\tau[\lambda^*_{\rm POLY3}]$.
(iii) Third, we benchmark against a brute-force optimal protocol (BF) $\lambda^*_{\rm BF}(t)$, taken to be piecewise linear with parameters $\{\lambda_{\rm BF}(n\delta)\}_{n=0}^{\tau/\delta}$.
Although not guaranteed to be globally optimal, this is the most expressive ansatz we consider and is therefore treated as the numerical optimum, with the expectation that $W_\tau[\lambda^*_{\rm IMP3}] \simeq W_\tau[\lambda^*_{\rm BF}]$.
Because the number of parameters in $\lambda_{\rm BF}(t)$ grows with $\tau$, brute-force optimization becomes impractical at long durations, where thermodynamic geometry may provide an efficient alternative \cite{RL23}. We therefore use the short-time case as our most stringent benchmark and compute only $W_{\tau=0.5}[\lambda^*_{\rm BF}]$.

\begin{figure*}[t]
\includegraphics[keepaspectratio, scale=0.4]{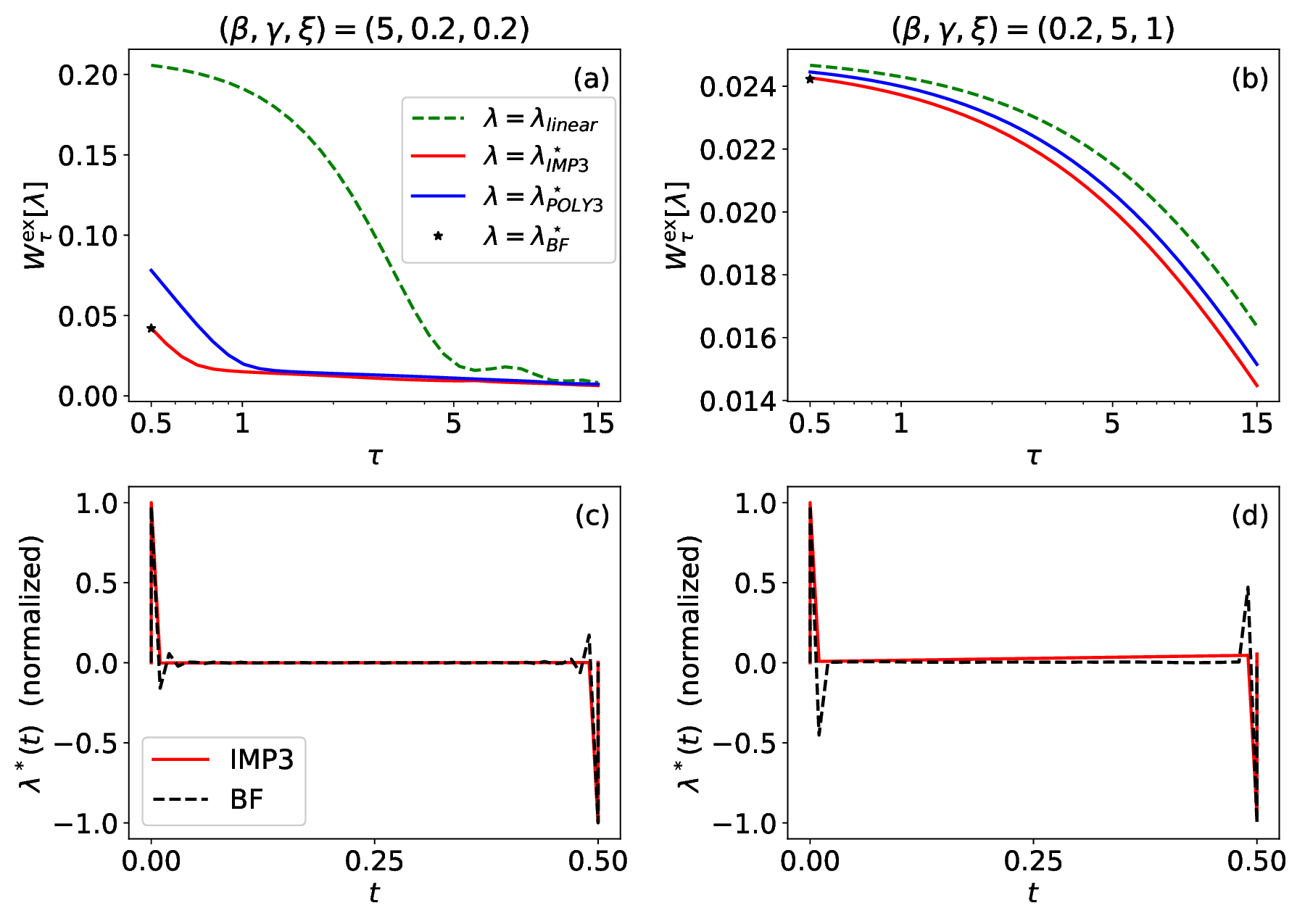}
\caption{
Excess work and optimal protocols for the driven two-level system with $(\beta,\gamma,\xi)=(5,0.2,0.2)$ (first column) and $(\beta,\gamma,\xi)=(0.2,5,1)$ (second column).
All quantities are in units of $\epsilon = \hbar = 1$.
(a,b) Excess work $W_{\tau}^{\rm ex}[\lambda]=W_\tau[\lambda]-\Delta F$ versus $\tau$ for the linear protocol (green dashed), IMP3 (red), POLY3 (blue), and BF (black stars).
(c,d) Normalized optimal protocols at $\tau=0.5$ for IMP3 (red) and BF (black dashed).
The vertical axis shows $\lambda^*(t)/\Lambda$ with $\Lambda=\max_{0\le t\le 0.5}|\lambda^*(t)|$; $\Lambda=313.5$ (IMP3) and $\Lambda=411.5$ (BF)
in panel (c) and $\Lambda=18.2$ (IMP3) and $\Lambda=127.2$ (BF) in panel (d).
}
\label{fig:Wdrive}
\end{figure*}

Our first example is a driven two-level system with $H_S(\lambda(t))=\epsilon \sigma_z/2 + \lambda(t)\sigma_x/2$ and $V_S=\sigma_x$, where $\epsilon$ is the level spacing and $\sigma_{x,y,z}$ are Pauli matrices. We set $\epsilon=1$, $\lambda_i=0$, and $\lambda_f=1$. For all 18 bath parameter sets, IMP3 consistently outperforms POLY3, yielding $W_\tau[\lambda^*_{\rm IMP3}] < W_\tau[\lambda^*_{\rm POLY3}]$ for $0.5\le \tau \le 15$, and closely matches the numerical optimum at $\tau = 0.5$, $W_{\tau=0.5}[\lambda^*_{\rm IMP3}] \simeq W_{\tau=0.5}[\lambda^*_{\rm BF}]$.
A key feature underlying this performance is the presence of impulse-like peaks in $\lambda^*_{\rm IMP3}$, even though the ansatz can represent smooth controls by tuning $h$.
Figures~\ref{fig:Wdrive}(a) and \ref{fig:Wdrive}(b) show the $\tau$ dependence of the excess work,
$W_\tau^{\rm ex}[\lambda]=W_\tau[\lambda]-\Delta F$, for two sets of bath parameters.
At $\tau=0.5$, the largest reduction achieved by IMP3 relative to the linear protocol occurs at $(\beta,\gamma,\xi)=(5,0.2,0.2)$ [\fref{fig:Wdrive}(a)].
To assess the attainability of the numerical optimum within IMP3, we estimate the relative error
$|(W_{\tau=0.5}[\lambda^*_{\rm IMP3}]-W_{\tau=0.5}[\lambda^*_{\rm BF}])/W_{\tau=0.5}[\lambda^*_{\rm BF}]|$.
The maximum error across all parameter sets is $6.2\%$, occurring at $(\beta,\gamma,\xi)=(0.2,5,1)$ [\fref{fig:Wdrive}(b)], where the IMP3 and brute-force results are visually indistinguishable.
The corresponding optimal protocols $\lambda^*_{\rm IMP3}(t)$, shown in \frefs{fig:Wdrive}(c) and \ref{fig:Wdrive}(d), exhibit pronounced impulse-like peaks.

\begin{figure}[t]
\includegraphics[keepaspectratio, scale=0.4]{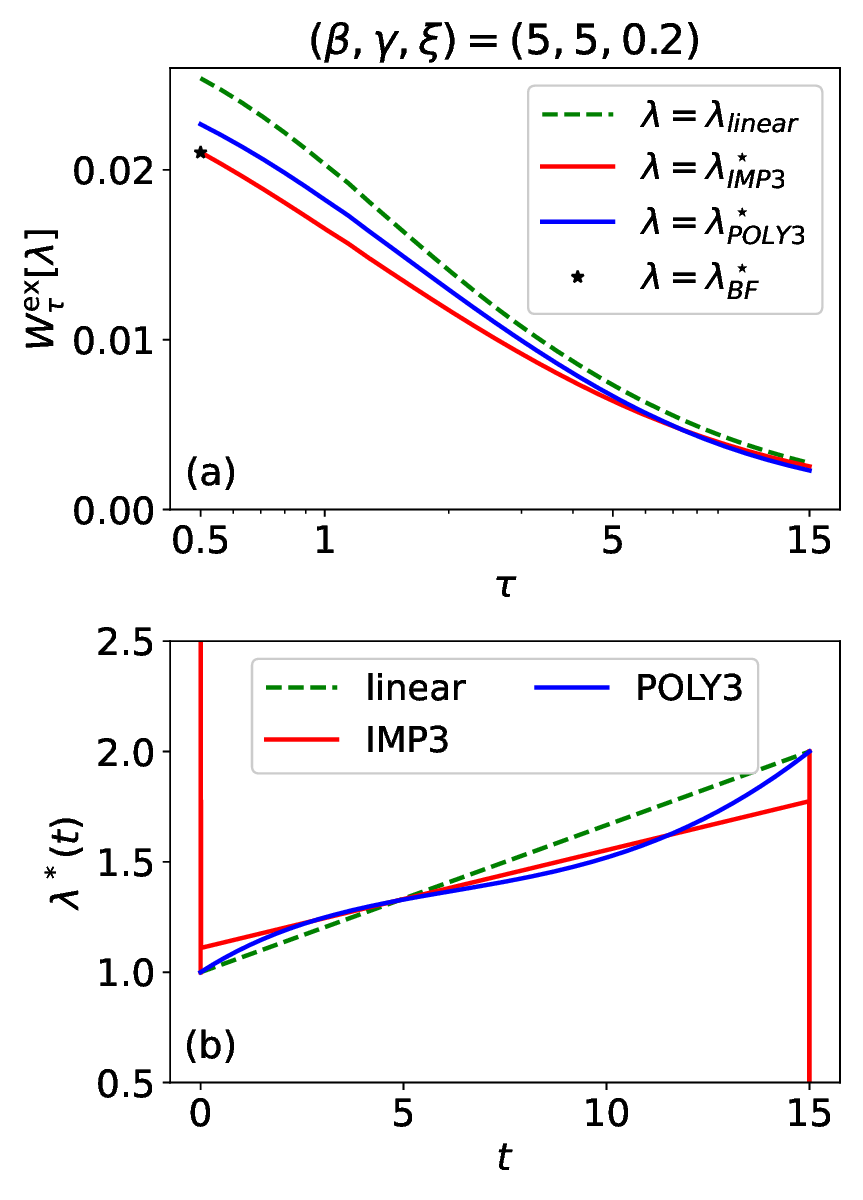}
\caption{
Excess work $W_{\tau}^{\rm ex}[\lambda]$ and optimal protocols for the tunable two-level system with $(\beta,\gamma,\xi)=(5,5,0.2)$. All quantities are in units of $\epsilon=\hbar=1$.
(a) $W_{\tau}^{\rm ex}[\lambda]$ versus $\tau$ for the linear protocol (green dashed), IMP3 (red), POLY3 (blue), and BF (black stars).
(b) Optimal protocols at $\tau=15$.
For IMP3, $\lambda^*_{\rm IMP3}(0^+) = 4.6$.
}
\label{fig:Wtunable}
\end{figure}

Our second example is a tunable two-level system with $H_S(\lambda(t))=\epsilon\,\lambda(t)\sigma_z/2$ and $V_S=\sigma_x$, where we set $\epsilon=1$, $\lambda_i=1$, and $\lambda_f=2$. As in the first example, we find $W_{\tau=0.5}[\lambda^*_{\rm IMP3}] \simeq W_{\tau=0.5}[\lambda^*_{\rm BF}]$ for all 18 bath parameter sets, confirming that impulse-like features are crucial for minimizing work at short times. The largest relative error is $0.10\%$, occurring at $(\beta,\gamma,\xi)=(0.2,5,1)$, where the deviation is negligible on the natural scale of the figure (not shown).

In contrast to the first example, at low temperature ($\beta=5$) we observe a slight but finite advantage of POLY3 over IMP3 at long durations ($\tau \gtrsim 5$).
This advantage is most pronounced at $(\beta,\gamma,\xi)=(5,5,0.2)$, where the $\tau$ dependence of the excess work is shown in \fref{fig:Wtunable}(a). In this case, $W_\tau[\lambda^*_{\rm POLY3}] < W_\tau[\lambda^*_{\rm IMP3}]$ for $\tau>7.4$. The corresponding optimal protocols at $\tau=15$ are plotted in \fref{fig:Wtunable}(b), with $\lambda^*_{\rm POLY3}(t)$ exhibiting pronounced nonlinearity.
These observations indicate a crossover: For $\tau \le 7.4$, impulse-like features near the boundaries efficiently reduce the work, whereas for $\tau>7.4$, nonlinear shaping in the intermediate region becomes more efficient.
Both behaviors can be captured within an extended impulse ansatz by introducing higher-order polynomials in the intermediate region (IMP$n$, $n\ge4$).

The significance of these results is twofold.
First, IMP3 offers an efficient route to estimating near-optimal work values.
With only three parameters, the optimization is substantially faster than the brute-force approach.
In our calculations, IMP3 typically converges within $10^2$ iterations for $0.5 \leq \tau \leq 15$, whereas the brute-force method requires on the order of $10^4$ iterations even at $\tau = 0.5$ with randomly generated initial parameters.
Second, IMP3 may provide an experimentally tractable optimization strategy.
While the brute-force optimal protocols, $\lambda^*_{\rm BF}(t)$, often exhibit multiple sign-flipping peaks near the boundaries [\frefs{fig:Wdrive}(c) and (d)], the close agreement in the excess work with the IMP3 values indicates that a single boundary peak captures the dominant contribution to work reduction.
Motivated by this observation, a minimal strategy is to fix $\alpha_1=\alpha_2=0$ and vary only $h$ to locate the minimum, similar in
spirit to \rcite{Loos24}.
We further note the possibility that the IMP3 parameters could be tuned directly in experiments using feedback-control techniques \cite{Biercuk09}.

The single-peak structure of IMP3 is motivated by the Markovian optimal solution and need not remain valid in non-Markovian settings.
Indeed, as shown in the Supplemental Material \cite{SM}, the numerically obtained optimal protocol for a Brownian particle in a moving harmonic trap with the same spectral density develops multiple sign-flipping impulses near the boundaries, similar to the brute-force solutions [\frefs{fig:Wdrive}(c) and (d)].
This demonstrates that non-Markovianity qualitatively modifies the optimal control.
Nevertheless, IMP3 with $\delta=0$ (corresponding to delta impulses) still yields an excellent approximation to the minimum work \cite{SM}.
This result indicates the presence of sub-optimal solutions and that they are efficiently captured by IMP3, highlighting its practical utility.

\begin{figure}[t]
\includegraphics[keepaspectratio, scale=0.4]{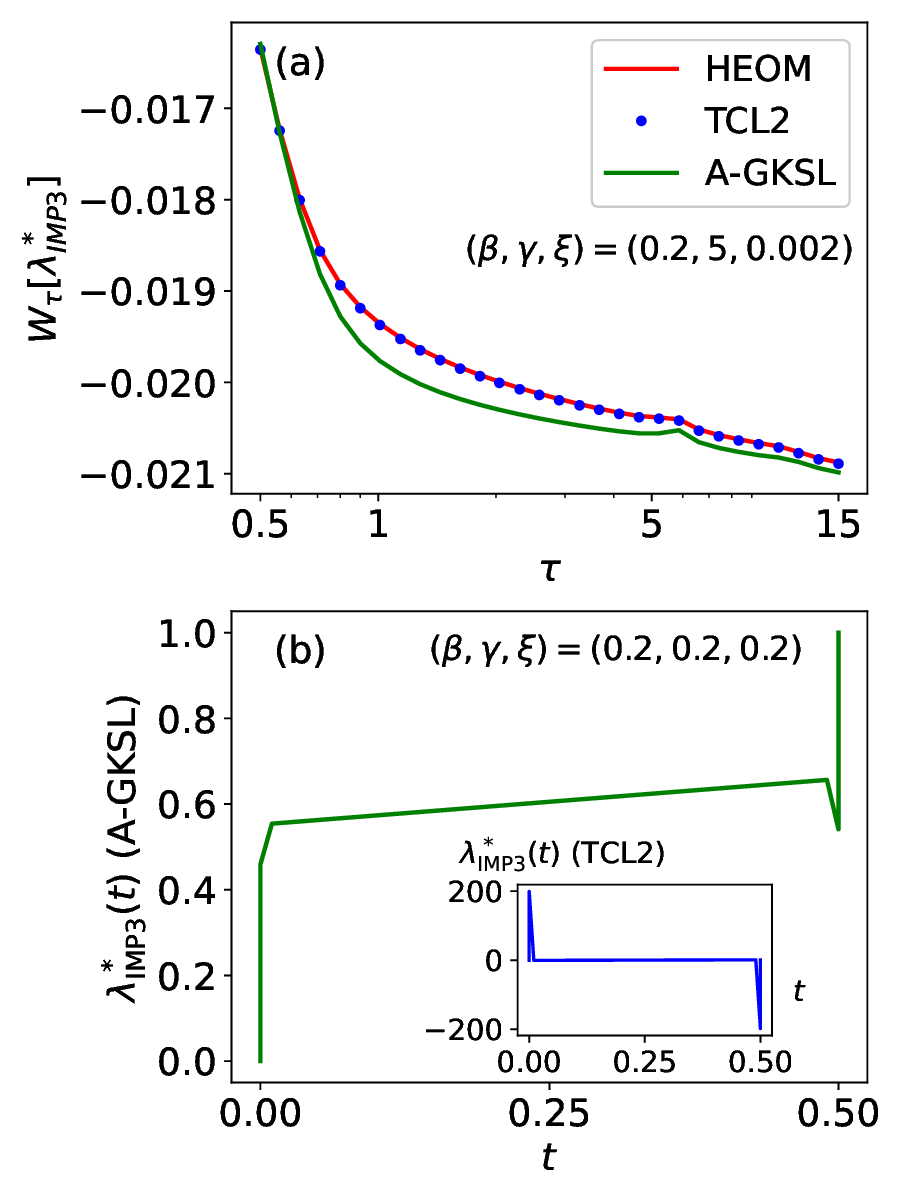}
\caption{
Comparison of methods for the driven two-level system. All quantities are in units of $\epsilon=\hbar=1$.
(a) Work as a function of $\tau$ for IMP3 obtained with HEOM (red), TCL2 (blue circles), and A-GKSL (green) for $(\beta,\gamma,\xi)=(0.2,5,0.002)$.
(b) IMP3 optimal protocol at $\tau=0.5$ obtained with A-GKSL for $(\beta,\gamma,\xi)=(0.2,0.2,0.2)$. The inset shows the corresponding TCL2 result.
}
\label{fig:weak_coupling}
\end{figure}

\sect{Caution in using GKSL master equations}
Quantum finite-time thermodynamics is typically analyzed within the weak-coupling approximation using a GKSL master equation.
Its derivation becomes subtle when the system Hamiltonian $H_S(t)$ is time dependent.
Here, we follow \rcite{Miller20} and employ the adiabatic Markovian master equation (A-GKSL) \cite{Childs01,Albash12,Yamaguchi17}.
Beyond the usual approximations for time-independent cases, A-GKSL additionally assumes that $H_S(t)$ varies slowly in time.
This assumption is questionable in our setting, where abrupt changes in $H_S(t)$, central to the optimal-work protocol, may invalidate the A-GKSL description.

We assess the validity of A-GKSL by applying it to the driven two-level system of the previous section.
For reference, we also consider the second-order time-convolutionless master equation (TCL2) \cite{Tokuyama76,Hashitsume77,Shibata77,Hanggi77,Breuer02,Ulrich04}, which is likewise based on the weak-coupling approximation.
Although TCL2 lacks a formal positivity guarantee, violations are not expected when the weak-coupling assumption is well satisfied \cite{Whitney08,Hartmann20}.
Since A-GKSL is derived from TCL2 by imposing additional approximations \cite{Albash12,Yamaguchi17}, discrepancies between the two signal the breakdown of these steps, most notably the assumption of a slowly varying $H_S(t)$.
Figure~\ref{fig:weak_coupling}(a) compares the optimal IMP3 predictions with $(\beta,\gamma,\xi)=(0.2,5,0.002)$.
Here, TCL2 closely coincides with HEOM, a numerically exact method, thereby confirming the validity of the weak-coupling approximation.
Nevertheless, A-GKSL departs noticeably around $\tau \simeq 1$.
For $(\beta,\gamma,\xi)=(0.2,0.2,0.2)$, the work values
$W_{\tau=0.5}[\lambda^*_{\rm IMP3}] = -1.14\times10^{-2}$ (HEOM), $-1.15\times10^{-2}$ (TCL2), and $-4.11\times10^{-3}$ (A-GKSL) again show that A-GKSL fails even when TCL2 tracks HEOM.
A more pronounced discrepancy arises in the optimal protocol: Figure~\ref{fig:weak_coupling}(b) shows that $\lambda^*_{\rm IMP3}(t)$ from A-GKSL exhibits only small boundary jumps, in sharp contrast to the impulse-like peaks predicted by TCL2 (inset), consistent with the HEOM behavior (not shown).

As emphasized in \rcite{Yamaguchi17}, A-GKSL can remain accurate even outside the adiabatic regime when the dynamics are dominated by the coherent part of the evolution.
Indeed, for some parameters, e.g. $(\beta,\gamma,\xi) = (5,0.2,0.2)$, A-GKSL reproduces the HEOM behavior despite the presence of strong impulses in the optimal protocol.
However, the preceding examples caution against using A-GKSL in optimization problems, as its underlying approximations may fail for certain choices of the optimization variable.
In the present system, TCL2 exhibits more robust performance across parameter ranges, and we therefore recommend its use, rather than A-GKSL, when analyzing the weak-coupling regime.

\sect{Conclusion}
In summary, we employed the bath oscillator model to study finite-time thermodynamics in the fully quantum regime, without invoking overdamped or Markovian limits.
Inspired by previous insights on boundary discontinuities, we introduced IMP3 (\fref{fig:ansatz}), a minimal ansatz that incorporates impulse-like features at the beginning and end times.
For representative two-level systems, IMP3 reproduces brute-force optimal results in the short-time regime, demonstrating its practical usefulness.
The impulse ansatz, possibly augmented by nonlinear terms in the intermediate region, is readily applicable to more complex quantum systems, and we leave such extensions for future investigation.

\sect{Acknowledgements}
I thank Yoshitaka Tanimura and Shoki Koyanagi for their valuable input throughout the research. This work was supported by JSPS KAKENHI Grant Number JP23KJ1157.

\sect{Data Availability}
The data that support the findings of this article are openly available \cite{Data}.


\begin{thebibliography}{66}
\bibitem{dscr.vs.cont}
    The proof in \rcite{Tasaki02} is for a Hamiltonian with discrete spectrum and the applicability to the current Hamiltonian with continuously distributed oscillator modes is an open question.
\bibitem{OhmicAnomaly} We avoid using the Ohmic spectral density, $J(\omega) \propto \omega$, which reproduces the Markovian Langevin equation in the classical limit, for two reasons. First, the contribution of the Matsubara terms to the noise correlation diverges, and it yields anomalous behavior (see, e.g., Sec.~7 of \rcite{GSI88} for the harmonic oscillator case).
    Second is to test IMP3, inspired by the solution to a Markovian case, againt more challenging non-Markovian cases.
\end{thebibliography}
\end{document}


\title{Supplemental Material for ``Work-minimizing protocols in driven-dissipative quantum systems: An impulse-ansatz approach''}


\author{Masaaki Tokieda}
\email{tokieda.masaaki.b01@kyoto-u.jp}
\affiliation{Department of Chemistry, Graduate School of Science, Kyoto University, Kyoto, Japan}

\date{\today}

\maketitle

\tableofcontents

\section{Optimal protocols for a Brownian particle in a moving harmonic trap}
\label{app:MovingTrap}

In this section, we consider a Brownian particle confined in a harmonic trap with a controllable center.
In the Markovian limit, the exact optimal protocol is known in closed form in both underdamped and overdamped regimes.
For non-Markovian dynamics, we determine the optimal protocol numerically and assess the performance of the three-parameter impulse ansatz (IMP3).

\subsection{Problem settings}

Consider the bath oscillator model
\begin{equation}
    H_S(\lambda(t)) = \epsilon \, a^\dagger \! a + \frac{\lambda(t)}{2} (a + a^\dagger) + H_{\rm CT}, \ \ \ \ \ V_S  = a + a^\dagger,
    \label{eq:MovingTrap_HO}
\end{equation}
where $a$ ($a^\dagger$) is the annihilation (creation) operator,
$\epsilon$ is the oscillator frequency, and $H_{\rm CT} = V_S^2 \sum_m (c_m^2/\omega_m)$ is the counter term to remove the bath-induced potential renormalization.
The work $W_\tau[\lambda]$ is given by
\begin{equation}
    W_\tau [\lambda] = \frac{1}{\sqrt{2}} \int_0^\tau dt \, \dot{\lambda}(t) \braket{q}_t,
    \label{eq:MovingTrap_W}
\end{equation}
with $q = (a + a^\dagger) / \sqrt{2}$ and $\braket{q}_t = {\rm tr}_S[q \, \rho_S(t)]$.
For an initial thermal state $\rho(0) \propto e^{- \beta H(\lambda_i)}$, $\braket{q}_t$ obeys
\begin{equation}
    \ddot{\braket{q}}_t
    = - 2 \epsilon \int_0^t ds \, \Delta (t-s) \dot{\braket{q}}_s
    - \epsilon^2 \braket{q}_t
    - \frac{\epsilon \lambda(t)}{\sqrt{2}},
    \label{eq:MovingTrap_EOM}
\end{equation}
with the friction kernel $\Delta(t) = (2/\pi) \int_{0}^{\infty} \, d\omega  (J(\omega)/\omega) \cos(\omega t)$ and initial conditions
\begin{equation}
    \braket{q}_0 = - \frac{\lambda_i}{\sqrt{2} \epsilon}, \ \ \ \ \ \dot{\braket{q}}_0 = 0.
    \label{eq:MovingTrap_initial.condition}
\end{equation}

We make two remarks.
First, for this linear system, there is no distinction between the classical and quantum treatments.
Second, the optimal protocol $\lambda^*(t)$ and the corresponding trajectory $\braket{q}_t$ exhibit time-reversal symmetry \cite{Loos24}.
Specifically, when $\lambda_i = 0$, $\lambda^*(t)$ satisfies
\begin{equation}
    \lambda^*(t) = \lambda_f - \lambda^*(\tau-t).
    \label{eq:MovingTrap_symmetry}
\end{equation}

\subsection{Ohmic spectral density: Markovian limit}

For the Ohmic spectral density
\begin{equation}
    J_{\rm Ohm}(\omega) = \frac{\zeta}{2 \epsilon} \omega,
    \label{eq:MovingTrap_Johm}
\end{equation}
we find $\Delta (t) = (\zeta/\epsilon) \delta(t)$, which yields
\begin{equation}
    \ddot{\braket{q}}_t
    = - \zeta \dot{\braket{q}}_t
    - \epsilon^2 \braket{q}_t
    - \frac{\epsilon \lambda(t)}{\sqrt{2}}.
    \label{eq:MovingTrap_ohm.EOM}
\end{equation}
In this Markovian limit, the optimal protocol $\lambda^*(t)$ minimizing  \eref{eq:MovingTrap_W} under \erefs{eq:MovingTrap_initial.condition} and (\ref{eq:MovingTrap_ohm.EOM}) was derived in \rcite{MSS08} and given by
\begin{equation}
    \lambda^*(0<t<\tau) = \lambda_i + \frac{1+\epsilon^2 t / \zeta}{2 + \epsilon^2 \tau / \zeta}(\lambda_f-\lambda_i)
    + \frac{2 (\lambda_f-\lambda_i) / \zeta}{2 + \epsilon^2 \tau  / \zeta} (\delta(t) - \delta(t-\tau)).
    \label{eq:MovingTrap_opt}
\end{equation}
It features delta-functional impulses at $t=0,\tau$ together with discontinuous jumps.

In the overdamped limit, \eref{eq:MovingTrap_ohm.EOM} reduces to
\begin{equation}
    \zeta \dot{\braket{q}}_t
    = - \epsilon^2 \braket{q}_t
    - \frac{\epsilon \lambda(t)}{\sqrt{2}},
    \label{eq:MovingTrap_ohm.EOM.overdamped}
\end{equation}
and the optimal protocol becomes \cite{SS07}
\begin{equation}
    \lambda^*(0<t<\tau) = \lambda_i + \frac{1+\epsilon^2 t / \zeta}{2 + \epsilon^2 \tau / \zeta}(\lambda_f-\lambda_i).
    \label{eq:MovingTrap_opt.overdamped}
\end{equation}
This matches \eref{eq:MovingTrap_opt} but without the impulse terms.

\subsection{General spectral density}

Here, we extend the analysis to more general spectral densities that generate non-Markovian dynamics.
For clarity, we set $\lambda_i = 0$ throughout this section.

\subsubsection{Numerical optimal protocol}

In the underdamped regime [\eref{eq:MovingTrap_EOM}], no analytic solution for the optimal control field is known for a general spectral density to our knowledge.
To obtain the optimal protocol numerically, we first recast the problem as a convex optimization.
Using the Laplace-transform solution of \eref{eq:MovingTrap_EOM},
\begin{equation*}
    \braket{q}_t = \epsilon \int_0^t ds \, G_+(t-s) x(s),
\end{equation*}
with $x(t) = - \lambda(t)/\sqrt{2}$ and $G_+(t)$ defined by $\ddot{G}_+(t) = - 2 \epsilon \int_0^t ds \, \Delta(t-s) \dot{G}_+(s) - \epsilon^2 G_+(t)$, $G_+(0) = 0$, $\dot{G}_+(0) = 1$,
the work [\eref{eq:MovingTrap_W}] becomes
\begin{equation}
    W_\tau[x] = \int_0^\tau dt \int_0^t ds \, A(t-s) x(t) x(s) - \int_0^\tau dt \, b(t)x(t),
    \label{eq:MovingTrap_Wconvex}
\end{equation}
with $A(t-s) = \epsilon \dot{G}_+(t-s)$ and $b(t) = x(\tau) \epsilon G_+(\tau-t)$.
This expression is quadratic in $x(t)$, allowing the optimal solution to be obtained straightforwardly.

\begin{figure}[t!]
\includegraphics[keepaspectratio, scale=0.5]{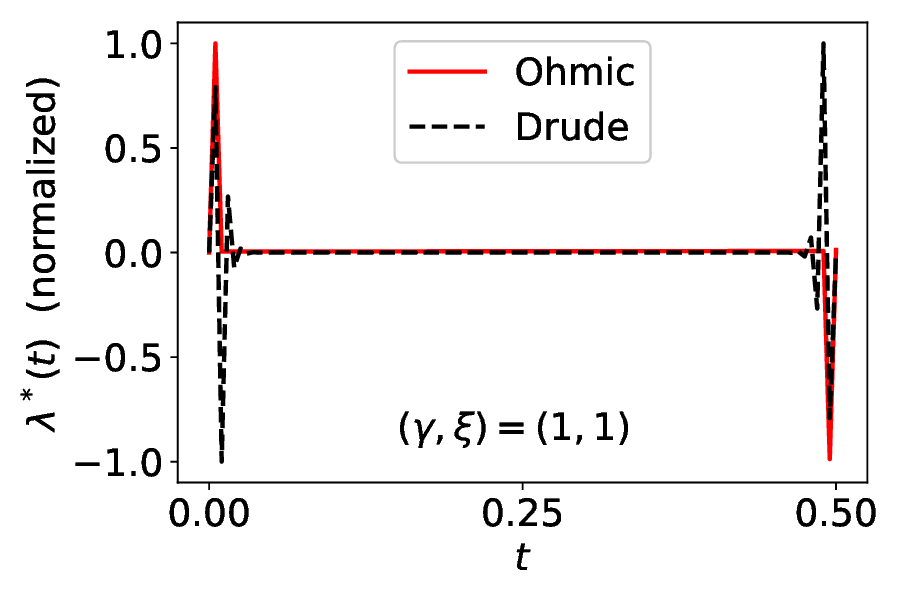}
\caption{
Normalized optimal protocols at $\tau = 0.5$ for the Ohmic (red) and Drude (dashed black) spectral densities.
All dimensional quantities are in units of $\epsilon = \hbar = 1$.
The vertical axis shows $\lambda^*(t)/\Lambda$, where $\Lambda=\max_{0\le t\le 0.5}|\lambda^*(t)|$, with $\Lambda = 81$ (Ohmic) and $\Lambda = 9.9\times 10^3$ (Drude).
}
\label{fig:opt_ohm_drude}
\end{figure}

As an illustration, we consider the Drude spectral density $J(\omega) = \gamma^2 \xi \omega / (\omega^2 + \gamma^2)$, for which $G_+(t)$ can be evaluated from $\Delta(t) = \gamma \xi e^{-\gamma |t|}$.
The discretized protocol values $\{ \lambda (n\delta) \}_{n=1}^{\tau/\delta-1}$ are treated as optimization parameters for minimizing $W_\tau$, subject to the time-reversal symmetry constraint \eref{eq:MovingTrap_symmetry}.
The integrals in \eref{eq:MovingTrap_Wconvex} are evaluated using the trapezoidal rule.
For $\tau = 0.5$, $\delta = 5 \times 10^{-3}$, $\epsilon = 1$, $(\lambda_i, \lambda_f) = (0,1)$, and $(\gamma, \xi) = (1, 1)$, the resulting optimal protocol $\lambda^*(t)$ is shown in \fref{fig:opt_ohm_drude} (dashed black).
For comparison, the Ohmic result \eref{eq:MovingTrap_Johm} with $\zeta=1$ is shown in red, which exhibits sharp impulse-like peaks consistent with the analytic solution \eref{eq:MovingTrap_opt}, including the peak height, the linear-region slope and intercept, and the value of $W_\tau[\lambda^*]$.
In contrast, the Drude case develops multiple sign-flipping peaks near the boundaries, for which we currently lack an analytical explanation.

In the overdamped limit,
\begin{equation}
    2 \int_0^t ds \, \Delta (t-s) \dot{\braket{q}}_s =
    - \epsilon \braket{q}_t
    + x(t),
\label{eq:MovingTrap_overdamped}
\end{equation}
an analytical optimal protocol was obtained in \rcite{Loos24}.
The corresponding spectral density combines the Ohmic and Drude forms, $J(\omega) = \zeta \omega / (2 \epsilon) + \gamma^2 \xi \omega / (\omega^2 + \gamma^2)$, yielding $\Delta(t) = (\zeta/\epsilon) \delta(t) + \xi \gamma e^{-\gamma |t|}$.
Indeed, integrating by parts and introducing $\braket{q_b}_t = \gamma \int_0^t ds \, e^{-\gamma(t-s)} \braket{q}_s$, we obtain
\begin{eqnarray*}
    \frac{\zeta}{2 \xi \gamma \epsilon} \dot{\braket{q}}_t
    &=& - \frac{\epsilon}{2 \xi \gamma} \left[ \braket{q}_t - \frac{x(t)}{\epsilon} \right] - \left[ \braket{q}_t - \braket{q_b}_t \right], \\
    \frac{1}{\gamma} \dot{\braket{q_b}}_t &=& - [\braket{q_b}_t - \braket{q}_t],
\end{eqnarray*}
which is essentially equivalent to Eqs.~(9) and (10) of \rcite{Loos24} after taking expectation values.
The general solution of \eref{eq:MovingTrap_overdamped} is given by
\begin{equation*}
    \braket{q}_t = \int_0^t ds \, F(t-s) x(s),
\end{equation*}
where the Laplace transform of $F(t)$, $\hat{F}(z) = \int_0^\infty dt \, e^{-zt} F(t)$, is defined as
$\hat{F}(z) = [\epsilon + 2 z \hat{\Delta}(z)]^{-1}$.
The work then takes the form \eref{eq:MovingTrap_Wconvex} with $A(t-s) = \dot{F}(t-s) + 2 F(0) \delta(t-s)$ and $b(t) = x(\tau) F(\tau-t)$.
For the combined Ohmic-Drude spectral density, we likewise obtain the optimal protocol and find that increasing the Drude contribution suppresses the boundary jumps and induces strong nonlinearity in the intermediate region, consistent with \rcite{Loos24}.

\subsubsection{Comparison with IMP3}

The optimal protocol becomes more intricate in the presence of memory, raising the question of how well IMP3 performs in this setting.
To examine this, we adopt the delta-functional impuse
\begin{equation*}
    \lambda_{\rm IMP3}(t) = \alpha_1 t/\tau + \alpha_2 + 2 m [\delta(t) - \delta(t-\tau)].
\end{equation*}
Because $\lambda_{\rm IMP3}(t)$ depends linearly on the parameters $(\alpha_1,\alpha_2,m)$, minimizing $W_\tau[\lambda_{\rm IMP3}] = W_\tau(\alpha_1,\alpha_2,m)$ remains a convex optimization problem.
The optimal parameters $(\alpha_1^*,\alpha_2^*,m^*)$ follow from the stationarity conditions
\begin{equation*}
    \frac{\partial}{\partial x} W_\tau(\alpha_1^*, \alpha_2^*, m^*) = 0 \ \ (x = \alpha_1, \alpha_2, m).
\end{equation*}

\begin{figure}[t!]
\includegraphics[keepaspectratio, scale=0.5]{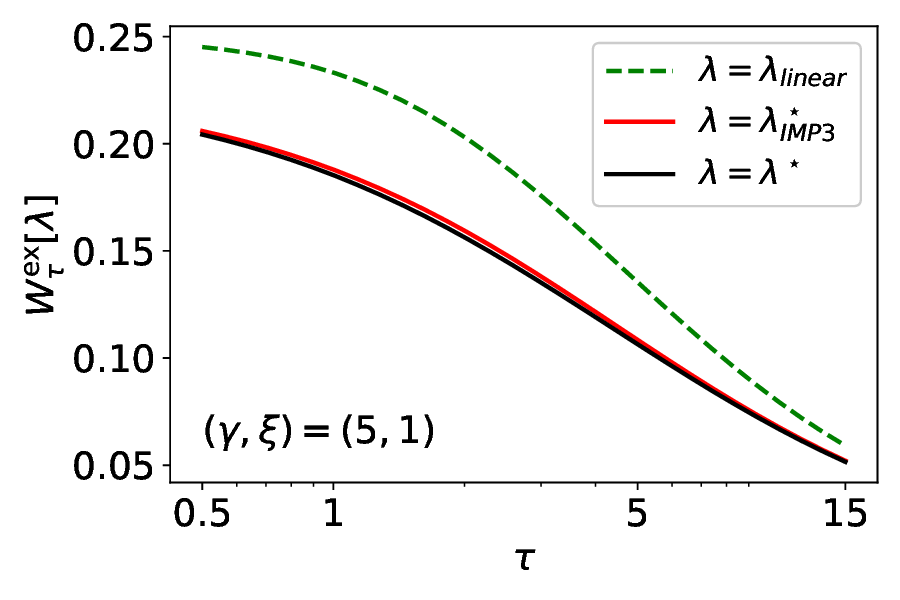}
\caption{
Excess work as a function of $\tau$ for the linear (green dashed) and IMP3 (red) protocols. The black curve indicates the global optimum.
All quantities are in units of $\epsilon = \hbar = 1$.
}
\label{fig:MovingTrap_IMP3}
\end{figure}

We assess the performance of IMP3 by comparing $W_\tau[\lambda^*_{\rm IMP3}]$ with the global minimum $W_\tau[\lambda^*]$ obtained as in \fref{fig:opt_ohm_drude}.
We consider $0.5 \le \tau \le 15$, $\epsilon=1$, $(\lambda_i,\lambda_f)=(0,1)$, and the Drude spectral density with six parameter sets with $\gamma = 0.2,1,5$ and $\xi = 0.2,1$.
To ensure that $W_\tau[\lambda^*]$ is indeed the global optimum, we vary the discretization step and confirm that $\delta = 2\times10^{-3}$ yields converged results for all bath parameters and all $\tau$.
Figure~\ref{fig:MovingTrap_IMP3} shows the case $(\gamma,\xi)=(5,1)$.
IMP3 yields a substantial reduction in work relative to the linear protocol $\lambda_{\rm linear}(t)=(\lambda_f-\lambda_i)t/\tau+\lambda_i$.
This parameter set gives the largest deviation at $\tau=0.9$,
$|W_{\tau=0.9}[\lambda^*_{\rm IMP3}] - W_{\tau=0.9}[\lambda^*]|/|W_{\tau=0.9}[\lambda^*]| = 4.0\%$,
yet IMP3 agrees well with the global optimum across the full range $0.5 \le \tau \le 15$.
This result supports the applicability of IMP3 in non-Markovian regimes.

\section{Computational details}
\label{compt}

This appendix summarizes the computational procedures used in the main text.

\subsection{Methodologies}

The work $W_\tau[\lambda]$ can be evaluated given the reduced state $\rho_S(t)$.
To compute its evolution, we employ the three methods in the main text: The hierarchical equations of motion (HEOM), the second-order time-convolutionless master equation (TCL2), and the adiabatic Markovian master equation (A-GKSL).
For completeness, we briefly outline each method below.
Throughout this appendix, we assume an initially factorized state,
\begin{equation}
    \rho(0) = \rho_S(0) \otimes \frac{e^{-\beta H_B}}{{\rm tr}_B [e^{-\beta H_B}]}.
    \label{eq:compt_rho0}
\end{equation}

\subsubsection{HEOM}

With the initial state \eref{eq:compt_rho0}, the reduced state evolution can be expressed as \cite{FV63,IF09}
\begin{equation}
    \rho_S(t) = \mathcal{U}_S(t) \, \mathcal{T} [\mathcal{M}(t)] \, \rho_S(0),
    \label{eq:method_rhoS}
\end{equation}
with the propagator
\begin{equation}
    \mathcal{M}(t) =
    \exp \left[ - \int_0^t ds \int_0^s du  V_S^I(s)^\times
    \left\{ L(s-u) V_S^I(u)^L - \bar{L}(s-u) V_S^I(u)^R \right\} \right],
\label{eq:method_M}
\end{equation}
where we introduce the chronological time-ordering $\mathcal{T}$, the superoperator notations $o^\times \rho= [o, \rho]$, $o^L \rho= o \rho$, $o^R \rho = \rho o$ for any operators $\rho$ and $o$, the unitary transformation $\mathcal{U}_S(t) = \mathcal{T}\left[ \exp \left\{ -i \int_0^t ds \, H_S(\lambda(s))^\times \right\} \right]$, the interaction picture $V_S^I(t) = \mathcal{U}_S^\dagger(t) V_S$, and the bath correlation function $L(t)$ with $\bar{L}(t)$ denoting its complex conjugation.
We assume that the bath correlation function admits the exponential expansion
\begin{equation}
    L(t \geq 0) = \sum_{k=1}^K d_k e^{-z_k t} + 2 \eta  \delta(t),
    \label{eq:compt_L}
\end{equation}
with real $\eta$ and $\{ z_k \}_{k=1}^K$ closed under complex conjugation so that there exist $\{ d'_k \}_{k=1}^K$ satisfying $\overline{\left( \sum_{k=1}^K d_k e^{-z_k t}\right)} = \sum_{k=1}^K d'_k e^{-z_k t}$.
Using this form, the propagator $\mathcal{M}(t)$ can be expressed as
\begin{equation*}
    \mathcal{M}(t) =
    \exp \left[ - \eta \int_0^t ds \, V_S^I (s)^\times V_S^I (s)^\times - \int_0^t ds \,
    V_S^I (s)^\times \, \sum_{k=1}^K \mathcal{Y}_k(s) \right],
\end{equation*}
with
\begin{equation*}
    \mathcal{Y}_k(s) = \int_0^s du \,
    \left( d_k V_S^I(u)^L - d'_k V_S^I(u)^R \right) e^{-z_k (s-u)}.
\end{equation*}
The time derivative of $\mathcal{T}[\mathcal{M}(t)]$ is given by
\begin{equation*}
    \frac{d}{dt} \mathcal{T}[\mathcal{M}(t)]
    = - \eta V_S^I (t)^\times V_S^I (t)^\times \mathcal{T}[\mathcal{M}(t)]
    - V_S^I(t)^\times \, \mathcal{T}\left[ \sum_{k=1}^{K} \mathcal{Y}_k(t) \mathcal{M}(t) \right].
\end{equation*}
Since the latest time in $\mathcal{T}$ is $t$,  $V_S^I(t)^\times$ can be taken outside the time-ordering. In contrast, $\mathcal{Y}_k(t)$ includes superoperators evaluated at earlier times and must remain inside.
To obtain a closed set of equations, we introduce the auxiliary system operators
\begin{equation}
\begin{gathered}
    \rho_{\bm{j}} (t) = \mathcal{U}_S(t) \, \mathcal{T} \left[
    \prod_{k=1}^{K} \left\{ \frac{ \left\{ \mathcal{Y}_k (t) \right\}^{j_k} }{\sqrt{j_k!}} \right\} \mathcal{M}(t) \right] \rho_S(0),
\end{gathered}
\label{eq:compt_rhoj}
\end{equation}
with $\bm{j}^\top = [j_1, \dots, j_{K}]$ and $j_k \in \mathbb{Z}_{\geq 0} \ (k = 1, \dots, K)$.
The reduced density operator corresponds to the element with $\bm{j} = \bm{0}$: $\rho_{\bm{0}} (t) = \rho_S(t)$.
Taking the time derivative yields the HEOM:
\begin{equation}
\begin{gathered}
    \dot{\rho}_{\bm{j}} (t)
    = - \left[ i H_S(\lambda(t))^\times + \sum_{k=1}^K z_k j_k + \eta V_S^\times V_S^\times \right] \rho_{\bm{j}} (t)
    + \sum_{k=1}^{K} \sqrt{j_k} \left( d_k V_S^L - d'_k V_S^R \right) \rho_{\bm{j}-\bm{e}_k} (t)
    - \sum_{k=1}^{K} \sqrt{j_k+1} \, V_S^\times \rho_{\bm{j}+\bm{e}_k}.
\end{gathered}
\label{eq:compt_HEOM}
\end{equation}

Equation~(\ref{eq:compt_rhoj}) indicates the initial conditions $\rho_{\bm{0}}(0) = \rho_S(0)$ and $\rho_{\bm{j} \ne \bm{0}}(0) = 0$.
Solving \eref{eq:compt_HEOM} under these conditions yields the reduced density operator from the element with $\bm{j} = \bm{0}$.
Since \eref{eq:compt_HEOM} generates an infinite hierarchy, we truncate by imposing
$\rho_{\bm{j}}(t)=0$ when $\sum_k j_k>D_H$, with hierarchy depth $D_H$.
The coefficients in \eref{eq:compt_L} are obtained by fitting the exact bath correlation function.
The accuracy of this expansion is validated following \rcite{MT25}.

\subsubsection{TCL2}

Using the projection $\mathcal{P} X = {\rm tr}_S (X) \otimes \rho_B$, we obtain an evolution equation for $\rho_S(t)$ that is nonlocal in time, since $\dot{\rho}_S(t)$ depends on $\rho_S(s)$ at earlier times $s < t$ through a time-convolution integral.
This equation can be transformed into a formally exact time-local form, known as the time-convolutionless (TCL) master equation, although its exact expression is generally too complicated for practical use.

For weak system–bath coupling, the TCL generator can be expanded perturbatively.
Up to second order, we obtain
\begin{equation}
    \dot{\rho}_S(t) = - i [H_S(\lambda(t)), \rho_S(t)] + [V_S, Q_S(t) \rho_S(t) - \rho_S(t) Q^\dagger_S(t)].
    \label{eq:compt_Redfield}
\end{equation}
with $Q_S(t) = \int_0^t ds \, L(s) V_S^I(-s)$.
A common Markov approximation replaces $Q_S(t)$ with $Q_S(\infty)$, yielding the Redfield equation, which can violate positivity.
In contrast, \eref{eq:compt_Redfield} has been shown to preserve positivity for various examples \cite{Whitney08,Hartmann20}, and we therefore use the second-order TCL without the Markov approximation.
To solve \eref{eq:compt_Redfield}, we follow \rcite{Ulrich04}, where the bath correlation function is expressed using the same exponential expansion in \eref{eq:compt_L} as employed in the HEOM calculations.

\subsubsection{A-GKSL}

Various derivations of the GKSL equation for time-dependent Hamiltonians have been proposed.
Here, we use A-GKSL, which assumes that $H_S(t)$ varies slowly.
Starting from the Redfield equation, complete positivity is recovered by performing the secular approximation.
For time-dependent $H_S(t)$, the assumption of a slowly varying $H_S(t)$ is required to carry out the secular approximation in the instantaneous eigenbasis of $H_S(t)$ \cite{Yamaguchi17}.
Operationally, the master equation is obtained by first deriving the GKSL equation for a time-independent Hamiltonian and then promoting eigenvalues and eigenstates to their instantaneous counterparts.

The two-level systems considered in the main text can be expressed as
$H_S(t) = \Omega_t \sigma_z^{\theta_t} / 2$, where $\sigma_i^{\theta} = U_\theta \sigma_i U_\theta^\dagger \ (i=x,y,z)$ with $U_\theta = [\cos(\theta/2) \ -\sin(\theta/2); \ \sin(\theta/2) \ \cos(\theta/2)]$, $0 \leq \theta_t < 2 \pi$,
where $[\Omega_t,\tan (\theta_t), {\rm sgn}(\cos(\theta_t))] = [\sqrt{\epsilon^2 + \lambda(t)^2}, \lambda(t)/\epsilon,+]$ for the driven system and $[\Omega_t,\tan (\theta_t), {\rm sgn}(\cos(\theta_t))] = [\epsilon|\lambda(t)|, 0,{\rm sgn}(\lambda(t))]$ for the tunable system.
In this case, A-GKSL is given by
\begin{equation*}
\begin{gathered}
    \dot{\rho}_S(t) = -i \left[ \left( \Omega_t + 2 \cos^2(\theta_t) {\rm Im}[d(\Omega_t)]\right) \sigma_z^{\theta_t}/2, \rho_S(t) \right] \\
    + 2 \cos^2(\theta_t) J(\Omega_t) \left\{ (1 + n_\beta(\Omega_t)) \mathcal{D}[\sigma_-^{\theta_t}] \rho_S(t)
    + n_\beta(\Omega_t) \mathcal{D}[\sigma_+^{\theta_t}] \rho_S(t) \right\}
    + 2 \sin^2(\theta_t) d(0) \mathcal{D}[\sigma_z^{\theta_t}] \rho_S(t),
\end{gathered}
\end{equation*}
with $\sigma_{\pm}^{\theta} = (\sigma_x^\theta \pm i \sigma_y^\theta) / 2$,
$n_\beta(\omega) = (e^{\beta \omega} - 1)^{-1}$,
and $d(\omega) = \int_0^\infty dt \, {\rm Re} [L(t)] e^{i \omega t}$ with ${\rm Re}$ and ${\rm Im}$ denoting the real and imaginary parts, respectively.
The quantity $d(\Omega_t)$ is evaluated using
the same exponential representation of $L(t)$ in \eref{eq:compt_L} as in the HEOM calculations.

\subsection{Evaluation and optimization of $W_\tau[\lambda]$}

To ensure that the initial state of the process is the equilibrium state of $H(\lambda_i)$, each master equation is first evolved with fixed $\lambda = \lambda_i$ until the steady state is reached.
Resetting this moment to $t = 0$, we then evolve the system with time dependent $\lambda(t)$ up to $t = \tau$.
All equations are integrated using the fourth-order Runge-Kutta method with a time step of $10^{-3}$, except for the tunable system with $\gamma=0.2$, where $2 \times 10^{-4}$ is used.
Convergence is verified by repeating the calculation with a ten-times smaller time step (and, for HEOM, increasing the hierarchy depth $D_H$ by 2) for both the linear protocol $\lambda_{\rm linear}$ and the initial guess.

The work is evaluated from the obtained $\rho_S(t)$ as
$W_\tau [\lambda] = {\rm tr}_S [H_S(\lambda_f) \rho_S(\tau) - H_S(\lambda_i) \rho_S(0)] - \int_0^\tau dt \, {\rm tr}_S \left[ H_S(\lambda(t)) \dot{\rho}_S(t) \right]$ with the integral computed using Simpson's rule on the same time grid.
For a parameterized protocol $\lambda(t)$, we minimize $W_\tau[\lambda]$ using the L-BFGS-B algorithm implemented in \texttt{scipy.optimize.minimize}.
The convergence tolerances are set to $10^{-10}$ both for the relative change in the objective function (\texttt{ftol}) and for the maximum norm of the gradient (\texttt{gtol}).
Applying this procedure to \eref{eq:MovingTrap_Wconvex}, we obtain an optimal protocol that closely agrees with the exact solution obtained by solving the corresponding convex optimization problem.

The free-energy difference $\Delta F$ is obtained using the linear protocol at sufficiently large $\tau$:
$\Delta F = W_{\tau = 2\times 10^4}[\lambda_{\rm linear}]$ for the driven system and
$\Delta F = W_{\tau = 2\times 10^3}[\lambda_{\rm linear}]$ for the tunable system.

\subsection{Initial guesses}

The L-BFGS-B algorithm requires an initial guess for the optimization variables.
We summarize the choices used in the main text.

\subsubsection{POLY3 ($\alpha_1, \alpha_2, \alpha_3$)}

For POLY3, we initialize the search with the linear protocol, $\alpha_1=\alpha_2=\alpha_3=0$.

\subsubsection{IMP3 ($h, \alpha_1, \alpha_2$)}

For the driven system, we construct the initial guess by exploiting its formal similarity to the Brownian oscillator in \eref{eq:MovingTrap_HO}.
Indeed, the system Hamiltonian can be expressed as $H_S(\lambda(t))=\epsilon c^\dagger c + \lambda(t)(c+c^\dagger)/2$ and $V_S=c+c^\dagger$ with fermionic annihilation (creation) operators $c$ ($c^\dagger$).
Using the corresponding bosonic result for the Ohmic case [\eref{eq:MovingTrap_Johm}], we take its optimal protocol [\eref{eq:MovingTrap_opt}] as the initial parameter choice:
\begin{equation}
\alpha_1=
\frac{(\lambda_f-\lambda_i)\epsilon^2/\zeta}{2+\epsilon^2\tau/\zeta},
\qquad
\alpha_2=
\lambda_i+\frac{\lambda_f-\lambda_i}{2+\epsilon^2\tau/\zeta},
\qquad
h=\frac{1}{\delta}\,
\frac{2 (\lambda_f-\lambda_i)/\zeta}{2+\epsilon^2\tau/\zeta},
\label{eq:compt_imp3_driven_initial}
\end{equation}
where $h$ is chosen to preserve the impulse area for finite width $\delta$.
To generalize beyond the Ohmic case, we identify $\zeta = 2 J_{\rm Ohm}(\epsilon)$ and replace it by $\zeta = 2 J(\epsilon)$ for a general spectral density $J(\epsilon)$.

For the tunable system, we take the results in the high-temperature limit.
In this limit with an Ohmic bath,  $\langle\sigma_z\rangle_t = {\rm tr}_S[\sigma_z \rho_S(t)]$
obeys \cite{CL83FP}
\begin{equation*}
\dot{\langle\sigma_z\rangle}_t
= -\frac{2\zeta}{\beta\epsilon}\langle\sigma_z\rangle_t
     - \zeta\,\lambda(t),
\qquad
\langle\sigma_z\rangle_0 = -\frac{\beta\epsilon\lambda_i}{2},
\end{equation*}
and the work is $W_\tau = (\epsilon/2)\int_0^\tau dt\,\dot{\lambda}(t)\langle\sigma_z\rangle_t$.
These equations map to the overdamped Brownian case \eref{eq:MovingTrap_ohm.EOM.overdamped} under
$\langle q\rangle_t\!\to\!\langle\sigma_z\rangle_t$,
$\epsilon\!\to\!\sqrt{2}/(\beta\epsilon)$,
$\zeta\!\to\!1/(\beta\epsilon\zeta)$,
up to an overall scale of the work.
Using \eref{eq:MovingTrap_opt.overdamped} with
$\epsilon^2/\zeta \to 2\zeta/(\beta\epsilon)$, the initial guess becomes
\begin{equation}
\alpha_1=\frac{(\lambda_f-\lambda_i)\,2\zeta/(\beta\epsilon)}
               {2+2\zeta\tau/(\beta\epsilon)},
\qquad
\alpha_2=\lambda_i+
          \frac{\lambda_f-\lambda_i}{2+2\zeta\tau/(\beta\epsilon)},
\qquad
h=0.
\label{eq:compt_imp3_tunable_initial}
\end{equation}
We set $\zeta = 2 J(\epsilon)$ as above, and we use the effective temperature $\beta \to (2/\epsilon)\tanh(\beta\epsilon/2)$.

Instead of $h$, we used the rescaled optimization variable $h' = h / (h^{\rm ref}/\delta)$, where $h^{\rm ref}/\delta = 2 (\alpha_2-\lambda_i)/(\zeta \delta)$
is the reference impulse height defined by the initial guess.
With this definition, the initial value is $h'=1$ for the driven system and $h'=0$ for the tunable system.

For the tunable system at $(\beta,\gamma,\xi)=(1,0.2,0.2)$, the optimized results obtained with the initial linear protocol, $(\alpha_1,\alpha_2,h’)=(\lambda_f-\lambda_i,\lambda_i,0)$, are identified as $\lambda^*_{\rm IMP3}$, since it yields smaller work values than \eref{eq:compt_imp3_tunable_initial} for certain $\tau$.

\subsubsection{Brute-force appraoch ($\{ \lambda_{\rm BF}(n \delta) \}_{n=0,1,\dots,\tau/\delta}$)}

For the brute-force approach, we consider three types of initial guesses.
The first is the linear protocol.
The second consists of ten random initial guesses generated by uniform sampling, $\lambda_{\rm BF}(n\delta)\in[-10,10]$.

The third is motivated by the observation that the resulting optimal protocols from the first two choices often exhibit multiple sign-flipping peaks near the boundaries.
To capture this structure, we introduce a six-parameter ansatz describing two asymmetric boundary impulses, referred to as 2IMP6.
The parameters are $\lambda(0^+)$, $\lambda(\delta)$, $\lambda(\tau-\delta)$, $\lambda(\tau^-)$, and $\{\alpha_i\}_{i=1,2}$, where the protocol is linear in the intermediate region, $\lambda(2\delta \leq t \leq \tau-2\delta) = \alpha_1 (t/\tau) + \alpha_2$.
Using the initial guesses \erefs{eq:compt_imp3_driven_initial} and (\ref{eq:compt_imp3_tunable_initial}), we optimize the 2IMP6 parameters and take the resulting protocol $\lambda_{\rm 2IMP6}^*(t)$ as an initial guess.

Among all cases, the protocol yielding the minimum work is identified as the brute-force optimal solution.